\documentclass[aps,pre,reprint]{revtex4-2}

\draft 
\usepackage{amssymb,amsmath}
\usepackage{graphicx}
\usepackage{dcolumn}
\usepackage{bm}
\usepackage{color}
\usepackage{float}
\usepackage{hyperref}
\usepackage[version=4]{mhchem}
\usepackage{relsize}
\usepackage{comment}

\newcommand{\M}{{\cal M}}
\newcommand{\W}{{\cal W}}
\newcommand{\R}{{\cal R}}
\newcommand{\Q}{{\cal Q}}
\newcommand{\K}{{\cal K}}
\newcommand{\cO}{{\cal O}}

\begin{document}

\title{Spectrally-wide acoustic frequency combs generated using
oscillations of polydisperse gas bubble clusters in liquids}

\author{Bui Quoc Huy Nguyen and Ivan S.~Maksymov}
\email{imaksymov@swin.edu.au}
\affiliation{Optical Sciences Centre, Swinburne University of
  Technology, Hawthorn, Victoria 3122, Australia\looseness=-1} 

\author{Sergey A.~Suslov}
\affiliation{Department of Mathematics, Swinburne University of
  Technology, Hawthorn, Victoria 3122, Australia\looseness=-1} 


\begin{abstract}
Acoustic frequency combs leverage unique properties of the optical
frequency comb technology in high-precision measurements and
innovative sensing in optically inaccessible environments such as
under water, under ground or inside living organisms. Because acoustic
combs with wide spectra would be required for many of these 
applications but techniques of their generation have not
yet been developed, here we propose a new approach to the
creation of spectrally-wide acoustic combs using oscillations
of polydisperse gas bubble clusters in liquids. By means
of numerical simulations we demonstrate that clusters consisting
of bubbles with precisely controlled sizes can produce wide
acoustic spectra composed of equally-spaced coherent peaks. We show
that under typical experimental conditions bubble clusters remain
stable over time required for a reliable recording of comb signals. We
also demonstrate that the spectral composition of combs can be tuned
by adjusting the number and size of bubbles in a cluster.               
\end{abstract}

\maketitle 

\section{Introduction}
Frequency combs (FCs) are spectra containing equidistant coherent peaks.
Although mostly optical FCs have found widespread practical and
fundamental applications so far \cite{Pic19, For19}, 
in general FCs can be generated using waves other than light.
For example, a number of acoustic, phononic and acousto-optical
FC techniques have recently been introduced
\cite{Cao14, Xio16, Cao16, Gan17, Mak17, Gar18, Wu19, Mak19,
Gor20, Fri20, Mak21}. Among them, the acoustic frequency comb (AFC)
techniques stand out because they hold the promise to enable
ultrasensitive vibration detectors \cite{Kum14}, phonon lasers
\cite{Gru10, Bea10} and quantum computers \cite{Sta12}.
AFCs can also find applications in precision measurements
in diverse physical, chemical and biological systems
in conditions, where using light---and hence optical FCs---poses
technical and fundamental limitations. For example, this is the case
in under-water distance measurements \cite{Wu19} and also in some
biomedical imaging modalities \cite{Cao14, Gan17, Mak19}. 

In our previous work \cite{Mak21}, we have theoretically and 
experimentally demonstrated the possibility of generating AFCs
using oscillations of a cluster of gas bubbles in liquids
\cite{Bre95, Hwa00, Nas13}. We used low-pressure harmonic
ultrasound signals with the frequency that is an order of
magnitude higher than the natural frequency of the bubble
cluster \cite{Min33, Hwa00}. The interaction of ultrasound waves with
an oscillating bubble at the natural frequency of a cluster results
in the amplitude modulation of cluster's response and the appearance 
of sidebands around the harmonic and ultraharmonic peaks of the
driving ultrasound wave. We demonstrated that such sideband structures
can be used as AFCs.

However, as with other AFC generation techniques 
\cite{Gan17, Gar18}, in our experiments the number of
sideband peaks usable as an AFC is small. At present,
this restriction presents numerous technological
challenges that shape research efforts in the
field of FCs \cite{Pic19, For19}. For example, similarly to
optical FCs, for many applications the AFC spectrum
has to span over an octave of a bandwidth---that is, the
highest frequency in the comb spectrum has to be at least
twice the lowest frequency. Of course, the spectrum
of an AFC can be extended using one of the techniques developed,
for example, for broadening the spectra of opto-electronic 
FCs \cite{Zha19_1} such as supercontinuum generation
using nonlinear optical effects. (here, the adoption
of optical techniques in the acoustic domain is possible
because of the analogy between nonlinear optical processes
in photonic devices and nonlinear acoustic processes
in liquids containing gas bubbles \cite{Mak19}.) Furthermore,
our analysis reported in \cite{Mak21} demonstrates
that the number of peaks in a bubble-generated AFC and
their relative magnitude can be increased by simultaneously
decreasing the frequency and increasing the pressure of the
ultrasound wave driving bubble oscillations. 
   
In the current work, we suggest an alternative strategy for broadening
spectra of AFCs generated using oscillations of gas bubbles. We
theoretically investigate the use of polydisperse clusters consisting
of mm-sized bubbles with equilibrium radii $R_{n0}=R_{10}/n$,
where $R_{10}$ is the equilibrium radius of the largest bubble in the
cluster and $n=1,2,3,\ldots$ is the number of bubbles in the
cluster. Although clusters with other bubble size distributions could
be used in the proposed approach, the specific ratio of equilibrium
radii investigated in this paper allows generating AFCs with a
quasi-continuum of equally-spaced peaks, which is convincingly
demonstrated below by numerical simulations of clusters with $n=4$
bubbles. In line with our previous experiments \cite{Mak21}, in our
analysis we consider low-pressure ultrasound waves (up to 10\,kPa). We
show that the ultrasound frequency can be chosen in a wide spectral
range above the natural oscillation frequency of individual bubbles
in the cluster. Our calculations demonstrate that these relaxed
technical specifications can greatly facilitate the generation and
recording of stable AFC signals. This is because at low pressure
insonification bubble clusters exhibit a regular behaviour until the 
bubble dynamics becomes affected by their aggregation \cite{Mak21}.
Moreover, formation a bubble cluster with mm-range equilibrium radii
of about $2/n$\,mm is technologically straightforward and can be
accomplished using only a simple bubble generator equipped with a
customised air diffuser \cite{Mak21}.

It is noteworthy that stable gas bubble clusters called
bubble grapes have been previously generated
\cite{Kob79, Blake_book, Doi96, Rab11} using low-pressure ultrasound
waves. However, as discussed in Sec.~\ref{sec:sec1},
bubble grapes are formed when the sign of the secondary Bjerknes force
is reversed and the corresponding equilibrium state 
becomes stable \cite{Zab84}. In contrast, we propose to
generate AFCs in a regime, where the bubbles attract each
other but the magnitude of the secondary Bjerknes force
is very small due to the disparate natural bubble frequencies
and the frequency of the driving ultrasound wave. As a result, the
inter-bubble distance and the overall arrangement of bubbles
within a cluster appear to be sufficiently stable for the observation
of well-pronounced AFC spectra without applying any additional
cluster stabilisation procedures.     

\section{Background theory\label{sec:sec1}}
The accepted model of nonlinear oscillations of a single
gas bubble that does not undergo translational motion is given
by the Keller-Miksis (KM) equation \cite{Kel80} that takes into
account the decay of bubble oscillations due to viscous
dissipation and fluid compressibility. However, when the focus 
is on mm-sized gas bubbles oscillating at 20--100\,kHz frequencies
in water being driven by low-pressure ultrasound waves (with the
amplitude of up to 10\,kPa), the terms in the KM equation accounting
for acoustic losses become negligible \cite{Mak21}. Thus, the KM
equation effectively reduces to the classical Rayleigh-Plesset (RP)
equation \cite{Ray17, Ple49}.

As with a single bubble, the RP equation for a cluster consisting of
$N$ bubbles not undergoing translational motion can be obtained by
removing the acoustic losses terms from the generalised KM equation
for a bubble cluster \cite{Met97, Dza13}:
\begin{eqnarray} 
  R_n\frac{d^2R_n}{dt^2}
  &+&\frac{3}{2}\left(\frac{dR_n}{dt}\right)^2\nonumber\\ 
  &=&\frac{1}{\rho}\left(P\left(R_n,\frac{dR_n}{dt}\right)
      -P_\infty(t)\right)-P_{sn}\,,\label{eq:eq5}
\end{eqnarray}
where
\begin{eqnarray}
  P_n\left(R_n,\frac{dR_n}{dt}\right)
  &=&\left(P_0-P_v+\frac{2\sigma}{R_{n0}}\right)
      \left(\frac{R_{n0}}{R_n}\right)^{3\kappa}\nonumber\\
  &&-\frac{4\mu}{R_n}\frac{dR_n}{dt}-\frac{2\sigma}{R_n}\,.\label{eq:eq6}
\end{eqnarray}
The term accounting for the pressure acting on of the $n$th bubble
due to scattering of the incoming pressure wave by the neighbouring
bubbles in a cluster is given by
\begin{equation}
  \label{eq:eq6_1}
  P_{sn}=\sum_{l=1, l\neq n}^{N}\dfrac{1}{d_{nl}}\left( R_l^2\frac{d^2R_l}{dt^2}
    + 2R_l\left( \frac{dR_l}{dt}\right)^2 \right)\,,
\end{equation}
where $d_{nl}$ is the inter-bubble distance and $N$ is the total
number of bubbles in the cluster. The expression
$P_\infty(t)=P_0-P_v+\alpha\sin(\omega^*t)$ with the angular frequency
$\omega^*=2\pi f$ represents the periodically varied pressure in the 
liquid far from the bubble. Parameters $R_{n0}$, $R_n(t)$, $\mu$,
$\rho$, $\kappa$, $\sigma$, $\alpha$ and $f$ denote the equilibrium
and instantaneous radii of the $n$th bubble in the cluster, the
dynamic viscosity and the density of the liquid, the polytropic
exponent of a  gas entrapped in the bubble, the surface tension of a
gas-liquid interface and the amplitude and the frequency of a driving
ultrasound wave. Diffusion of the gas through the bubble surface is
neglected. 

When bubble oscillations are not affected by fluid compressibility,
which is the case in this work, the acoustic power scattered by the
$n$th bubble in the cluster in the far-field zone is \cite{Bre95}
\begin{equation}
  \label{eq:eq7}
  P_{scat}(R_n,t) = \frac{\rho R_n}{h}\left(R_n\ddot{R}_n+2\dot{R}_n^2\right)\,, 
\end{equation}
where $h$ is much larger than the spatial extent of the cluster.

Interactions of gas bubbles, including their radial oscillations
and translational motion driven by an acoustic pressure field,
has been a subject of intensive research (see, e.g.,
\cite{Nem83, Zab84, Wat93, Pel93, Doi95, Met97, Bar99, Har01,
Doi01, Mat05, Mac06, Met09, Yas10, Sad10, Jia12, Dza13, Lan15, Leb15};
for a review see \cite{Doinikov_book}). Most of these
works are based on the accepted models of spherical gas bubble oscillations
[Eqs.~(\ref{eq:eq5})--(\ref{eq:eq6_1})] and account for the action of Bjerknes
forces \cite{Bje06}. The primary Bjerknes force $F_{pB}$ is caused by
the acoustic pressure field \cite{Bje06, Lei90} while the secondary 
Bjerknes force $F_{sB}$ arises between two and more bubbles
in the same pressure field \cite{Doinikov_book}. The secondary Bjerknes
force between two gas bubbles is repulsive when the driving frequency
lies between the natural frequencies of the bubbles, otherwise it is
attractive \cite{Doinikov_book}. This theoretical prediction was
confirmed experimentally \cite{Kaz60, Cru75}. 

However, several important experimental observations cannot 
be explained using Bjerknes theory. These include the
formation of stable bubble grape clusters \cite{Kob79, Doi96}
and self-organisation effects in bubble-liquid mixtures \cite{Akh06}
that explain acoustic streaming phenomena \cite{Doi01, Pel04} and
underpin some techniques of bubble manipulation \cite{Lan15}. 

Kobelev and co-workers \cite{Kob79} were the first to report the
formation of bubble grapes as a byproduct of their experiment
targeting the attenuation of sound in liquids containing gas bubbles.
They demonstrated that nonlinear oscillations of gas bubbles
can not be responsible for the observed effect. The original
Bierknes force theory \cite{Bje06, Lei90, Doinikov_book} based on
the linear oscillations of gas bubbles is unable to explain this
phenomenon either since (i) it applies only to gas bubbles
separated by distances that are much larger then the bubble radii
and (ii) it only predicts whether the bubbles attract or repulse
depending on their natural frequencies. Thus, the only plausible
explanation of the formation of bubble grapes could be a reversal
in the secondary Bjerknes force from attractive to repulsive.

One of the first attempts to theoretically explain the formation of
bubble grapes was made by Nemtsov \cite{Nem83}. However,
his model did not account for wave scattering by
bubbles \cite{Doinikov_book}. Zabolotskaya \cite{Zab84}
was first to develop a model explaining the sign reversal of the
secondary Bjerknes force. Ozug and Prosperetti 
\cite{Ogu90} theoretically demonstrated the possibility of
sign reversal of the secondary Bjerknes force in the case of
nonlinear bubble oscillations driven by high-pressure acoustic waves.
However, the result presented in their work cannot explain 
linear processes underlying the formation of bubble grapes.    
Subsequently, Zabolotskaya's theory was extended by Doinikov
and Zavtrak \cite{Doinikov_book} and used to explain \cite{Doi96}
an intriguing observation of stable bubble structures in an
experiment involving strongly forced mm-sized gas bubbles
oscillating in low gravity conditions \cite{Blake_book}. 
An alternative interpretation of the sign reversal of the
secondary Bjerknes force was proposed in \cite{Ida03}.

However, although the experimental conditions in our work
reported in \cite{Mak21} indeed resemble those required for 
the formation of bubble grapes, the bubble clusters we observed
form due to a different mechanism that does not
involve the sign reversal of Bjerknes force. This 
is because we use mm-size bubble with the natural
frequencies of 1--3\,kHz but drive their oscillations 
with a low-pressure high-kHz-range ultrasound. As a result, 
although the aggregation and eventual coalescence of
bubble are inevitable in our experiments, they occur on a
timescale of several seconds and hence are mostly 
inconsequential for the generation of AFCs.  

\section{Interaction between two gas bubbles}
The interaction dynamics of gas bubbles oscillating in liquids
is very complex because the cluster geometry varies from experiment to
experiment and with time. Therefore, many theoretical works consider a
system of just two interacting bubbles surrounded by an idealised
liquid. This simplification allows reducing the complexity of
the model while accounting for the essential physics of bubble interaction. 

\subsection{Analysis of the RP equation for two interacting gas
  bubbles \label{sec:Poincar}}
To identify the main characteristics of nonlinear oscillations of
interacting gas bubbles relevant to the generation of AFCs, we conduct
an asymptotic analysis of Eq.~(\ref{eq:eq5}) extending our previous
model of nonlinear oscillations of a single gas bubble \cite{Mak21}.

We start with rewriting Eq.~(\ref{eq:eq5}) in the non-dimensional form 
using the equilibrium radius of the largest bubble in the cluster,
$R_{10}$, and $1/{\omega^*}$ as the length and time scales,
respectively, to introduce the non-dimensional quantities
$r_n={R_n(t)}/R_{10}$, $r_l={R_l(t)}/R_{10}$ and $\tau^* = \omega^*t$
\cite{Dza13}. Substituting these into Eq.~(\ref{eq:eq5}) we obtain
\begin{eqnarray}
  r_n {r_n}''&+&\frac{3}{2} {{r_n}'}^2
  =\left(\M+\frac{\W}{\Q_n}\right)\left(\frac{\Q_n}{r_n}\right)^\K
     -\frac{\W}{r_n}-\R\frac{{r_n}'}{r_n}\nonumber\\
  &-&\M-\M_e\sin\tau^*-\sum_{\substack{l=1\\l\neq n}}^N\zeta_{nl}
      \left(r_l^2r_l''+2r_l{r_l'}^2\right)\,,\label{eq:eq6_2}
\end{eqnarray}
where $\R=\dfrac{4\mu}{\rho\omega^*R^2_{10}}$,
$\W=\dfrac{2\sigma}{\rho\omega^{*2}R^3_{10}}$,
$\M=\dfrac{P_0-P_v}{\rho\omega^{*2}R^2_{10}}$,
$\M_e=\dfrac{\alpha}{\rho\omega^{*2}R^2_{10}}$,
$\zeta_{nl}=\dfrac{R_{10}}{d_{nl}}$,
$\K=3\kappa$ and $\Q_n=\dfrac{R_{n0}}{R_{10}}$.
Parameter $\M$ characterises elastic properties of the gas and its 
compressibility, $\W$ and $\R$ can be treated as inverse Weber and
Reynolds numbers, representing the surface tension and viscous
dissipation effects, respectively, and $\M_e$ is the measure of the
ultrasound forcing \cite{Sus12}. Parameters $\zeta_{nl}$ and $\Q_n$
are the inverse of the distance between the bubble centres and
the bubble radius relative to that of the largest bubble in the cluster,
respectively \cite{Dza13} and primes denote differentiation with
respect to $t$. As discussed in \cite{Sus12, Mak21}, $\K=4$ for
bubbles of sizes relevant to the AFC context and for the fluid
parameters, ultrasound pressure and frequency given in
Sec.~\ref{sec:Num} the maximum values of other parameters do not
exceed $\M=9.7\times10^{-4}$, $\W=7.4\times10^{-7}$,
$\R=6.5\times10^{-6}$ and $\M_e=9.9\times10^{-5}$. Therefore, the
effects of water viscosity and surface tention on bubble oscillations
are negligible and we set $\R=\W=0$ in what follows. Thus,
ultrasonically forced bubble oscillations can be assumed perfectly
periodic when the driving frequency is much higher than any
of the natural frequencies of the individual bubbles in the
cluster (i.e.~no resonances arises). This warrants
using a method similar to that in \cite{Mak21}.

We consider a cluster consisting of two gas bubbles with the 
non-dimensional equilibrium radii $r_{n0}=\Q_n$, $n=1,2$
($\Q_1\equiv1$). Following \cite{Che07, Mak21} we look for the
asymptotic solutions of Eq.~(\ref{eq:eq6_2}) in the form 
\begin{equation}
  r_n=\Q_n+\epsilon r_{n1}(\tau)+\epsilon^2r_{n2}(\tau)
  +\ldots\,,\quad n=1,2\,,\label{eq631}\\ 
\end{equation}
where $0<\epsilon\ll1$ is a parameter characterising the
amplitude of bubble oscillations used to distinguish between various
terms in the asymptotic series and $\tau=\omega\tau^*=\omega\omega^*t$.
At the first order of $\epsilon$ we obtain
\begin{eqnarray}
  &&\ddot r_{11}+\frac{\K\M}{\Q_1^2\omega^2}r_{11}
     +\frac{\Q_2^2}{\Q_1}\zeta_{12}\ddot{r}_{21}
     =\frac{p}{\Q_1}\sin(\Omega\tau)\,,\\
  &&\ddot r_{21}+\frac{\K\M}{\Q_2^2\omega^2}r_{21}
     +\frac{\Q_1^2}{\Q_2}\zeta_{12}\ddot{r}_{11}
     =\frac{p}{\Q_2}\sin(\Omega\tau)\,,
\end{eqnarray}
where overdots denote differentiation with respect to $\tau$ and we
write $(\M_e/\omega^2)\sin{\tau^*}\equiv-\epsilon p\sin(\Omega\tau)$ 
and introduce $\Omega\equiv1/\omega\gg1$. For convenience we also
choose $\omega^2=\K\M$, where $\K\M$ is Minnaert frequency
\cite{Min33, Mak21} of the largest bubble in the cluster. Finally, we 
obtain 
\begin{eqnarray}
  &&\ddot{r}_{11}+r_{11}+\Q_2^2\zeta_{12}\ddot{r}_{21}
     =p\sin(\Omega\tau)\,,\label{eqr11}\\
  &&\ddot{r}_{21}+\frac{1}{\Q_2^2}r_{21}+\frac{\zeta_{12}}{\Q_2}\ddot{r}_{11}
     =\frac{p}{\Q_2}\sin(\Omega\tau)\,.\label{eqr21}
\end{eqnarray}
At $\cO(\epsilon^2)$ equations then become:
\begin{eqnarray}
  \ddot{r}_{12}+r_{12}
  +\Q_2^2\zeta_{12}\ddot{r}_{22}&=&\frac{\K+1}{2}r_{11}^2
      -\frac{3}{2}\dot{r}_{11}^2-r_{11}\ddot{r}_{11} \nonumber\\
  &&-2\Q_2\zeta_{12}\left(\dot r_{21}^2+r_{21}\ddot r_{21}\right)\,,
      \label{eqr12}\\
  \ddot r_{22}+\frac{1}{\Q_2^2}r_{22}
  +\frac{\zeta_{12}}{\Q_2}\ddot r_{12}&=&\frac{\K+1}{\Q_2^2}r_{21}^2
      -\frac{3}{2\Q_2}\dot r_{21}^2-\frac{1}{\Q_2}{r}_{21}\ddot r_{21}
      \nonumber\\
  &&-2\frac{\zeta_{12}}{\Q_2}\left(\dot{r}_{11}^2+r_{11}\ddot r_{11}\right)
     \,.\label{eqr22}
\end{eqnarray}

Similarly to \cite{Mak21} we write the random initial conditions
as $r_{1}(0)=1+\epsilon a$, $r_{2}(0)=\Q_2+\epsilon b$,
$\dot{r_1}(0)=\epsilon c$ and $\dot{r_2}(0)=\epsilon d$ that results in 
\begin{equation}
  \label{eq:randIC}
  r_{11}(0)=a,\  r_{21}(0) =b,\ \dot{r}_{11}(0)=c,\ \dot{r}_{21}(0)=d\,.
\end{equation}
Subsequently, we obtain the leading order solutions
\begin{eqnarray}
  r_{11}(\tau)
  &=&C_1\cos{(\omega'_1\tau+\phi_1)} 
      + C_2\cos{(\omega'_2\tau+\phi_2)}\nonumber\\
  &&+B_1\sin{\Omega\tau}\,,\label{eq:1order1}\\
  r_{21}(\tau)
  &=&C_3\cos{(\omega'_1\tau+\phi_1)}
      +C_4\cos{(\omega'_2\tau+\phi_2)}\nonumber\\
  &&+B_2\sin{\Omega\tau}\,,\label{eq:1order2}
\end{eqnarray}
where
$\omega'_{1,2}=\dfrac{\sqrt{2}}{\sqrt{\Q_2^2+1
    \pm\sqrt{(\Q_2^2-1)^2+4\Q_2^3\zeta_{12}^2}}}$.
These frequencies depend on the inverse of the inter-bubble distance
$\zeta_{12}$, which is a well-established fact \cite{Zab84,
Doinikov_book, Man16}. Considering a particular case of
$\Q_2=\frac{1}{2}$, as expected, for non-interacting
distant bubbles with $\zeta_{12}\rightarrow0$ we obtain
$\omega'_1\to\omega'_{10}=1$ and $\omega'_2\to\omega'_{20}=2$.
In general, the leading order bubble response
will always contain three distinct frequencies: two bubble’s 
natural frequencies $\omega'_{1,2}$ and the driving ultrasound
frequency $\Omega$.

Coefficients $C_{1-4}$ and phase shifts $\phi_{1,2}$ in
Eqs~(\ref{eq:1order1}) and (\ref{eq:1order2}) depend on $\zeta_{12}$,
$\Omega$ and $p$. As shown in \cite{Mak21} their exact expressions can
be obtained for arbitrary initial conditions (\ref{eq:randIC}).
However, the resulting expressions are too long to be given here
explicitly and we only discuss the physical conclusions
that follow from them. Specifically, these coefficients
demonstrate that the spectra of both bubbles contain frequencies
$\Omega$ and $\omega'_{1,2}$. The magnitude of the $\omega_1'$ peak
is greater than that of $\omega_2'$ in the spectrum of bubble 1 and
vice versa. In the spectra of both bubbles, the amplitude of the peak
corresponding to the frequency of a neighbouring bubble decreases with
the distance between them and vanishes when the interaction between
them becomes negligible.

Analysis of Eqs~(\ref{eqr21}) and (\ref{eqr22}) can be performed
following  the procedure outlined in \cite{Mak21}. However, here we do
not pursuit this any further since for the purposes of the current work it
suffices to note that the right-hand sides of these equations contain
quadratic terms involving $r_{11}$ and $r_{12}$ and their derivatives.
Therefore, in addition to the harmonic components with frequencies
$\omega'_{1,2}$ solutions of Eqs~(\ref{eqr21}) and (\ref{eqr22}) will
include steady and periodic terms with frequencies equal to all possible
pair-wise sums and differences of $\omega'_{1,2}$ and $\Omega$:
$\omega'_{1,2}\pm\omega'_{2,1}$, $\Omega\pm\omega'_{1,2}$,
$2\omega_{1,2}'$ and $2\Omega$. 

\subsection{Bjerknes force between two oscillating
  bubbles \label{sec:Bjerknes}}
The asymptotic analysis in Sec.~\ref{sec:Poincar} shows that,
when the bubble oscillations are driven by a low-pressure ultrasound
wave, the magnitude of any nonlinear effects is proportional to
$\M_e^2\sim 9.8\times10^{-9}$ (as per experimental conditions 
in \cite{Mak21}). Hence, the nonlinearity is neglected in this section. 

Two physically equivalent dimensional expressions accounting
for a sign reversal of Bjerknes force arising between two oscillating gas
bubbles separated by a distance that is much larger than bubble radii
were derived previously in \cite{Zab84} and \cite{Doinikov_book}. From
the expression given by Eq.~(2.5) in \cite{Doinikov_book} we obtain
the leading term of the non-dimensional secondary Bjerknes 
force (scaled with $\rho{\omega^*}^2R_{10}^4$)
\begin{equation}
  F'_{sB}=-4\pi\zeta_{12}^2\Q_2^2\omega^2\epsilon^2\langle r_{11}\ddot
  r_{21}\rangle\,,\label{ndbjerk}
\end{equation}
where the angle brackets denote time averaging. Substituting
expressions (\ref{eq:1order1}) and (\ref{eq:1order2}) into
Eq.~(\ref{ndbjerk}), taking into account that
$\left|\omega_1'-\omega_2'\right|\sim1$ we obtain
\[
  F'_{sB}= 2\pi\zeta_{12}^2Q_2^2\omega^2\epsilon^2
  \left(B_1B_2\Omega^2+C_1C_3{\omega_1'}^2
    +C_2C_4{\omega_2'}^2\right)\,.
\]  
Evaluating coefficients $B_{1,2}$ and $C_{1-4}$ in the
limit $\zeta_{12}\to0$ finally leads to
\begin{equation}
  F'_{sB}=4\pi\zeta_{12}^2\Q_2^2\M_e^2
  \frac{\Omega^8}{(\Omega-1)^2(\Omega+1)^2}\,.\label{eq:eq4}
\end{equation}
Here we focus on a particular bubble oscillation regime \cite{Mak21}
characterised by  $\Omega\gg1$. In this case
\begin{equation}
  F'_{sB} \to 4\pi\zeta_{12}^2\Q_2^2\M_e^2\Omega^4=
  4\pi\zeta_{12}^2\Q_2^2\frac{\M_e^2}{\K^2\M^2}\,.
\end{equation}
By its definition, $Q_2<1$ and in the reference experiment
\cite{Mak21} $\M_e/(\K\M)\sim0.025$ and
$\zeta_{12}^2\lesssim0.04$. Therefore, we conclude that the secondary
Bjerknes force is small at the typical driving frequencies used in the
generation of bubble-based AFCs away from bubble resonances. This
provides an opportunity for measuring the acoustic bubble response and
recording the resulting signals for AFC applications before bubble
oscillations become affected by their aggregation. Our calculations
using a more rigorous model of a translational motion, the results of
which are presented in Sec.~\ref{sec:Num}, provide convincing
arguments in favour of this.

\subsection{Dynamics of multi-bubble clusters with translational
  motion} 
It has been shown in \cite{Doi04} that Eq.~(\ref{eq:eq5}) can be
extended to include the effect of a translational bubble motion.
The resulting system of differential equations reads
\begin{eqnarray}
  &&R_n\ddot{R}_n+\frac{3}{2}\dot{R}_n^2-\frac{P_n}{\rho} 
     =\frac{\dot{\bf{p}}_n}{4}-\sum\limits_{\substack{l=1\\l\neq n}}^N 
  \left\{\frac{R_l^2\ddot{R}_l+2R_l\dot{R}_l^2}{d_{nl}}\right.\nonumber\\
  &&\qquad+\frac{R_l^2}{2d_{nl}^3}({\bf p}_n-{\bf p}_l)\cdot (R_l\ddot{\bf p}_l 
     +\dot{R}_l\dot{\bf p}_n+5\dot{R}_l\dot{\bf p}_l)\nonumber\\ 
  &&\qquad-\frac{R_l^3}{4d_{nl}^3}\Big[\dot{\bf p}_l\cdot
     (\dot{\bf{p}}_n+2\dot{\bf p}_l)\nonumber\\
  &&\qquad+\frac{3}{d_{nl}^2}[\dot{\bf{p}}_l\cdot ({\bf{p}}_l-{\bf p}_n)] 
     [({\bf p}_n-{\bf p}_l)\cdot (\dot{\bf p}_n+2\dot{\bf p}_l)]\Big]\Bigg\}\,,
     \label{eq:eq8}\\
  &&\frac{1}{3}R_n\ddot{\bf{p}}_n+\dot{R}_n\dot{\bf{p}}_n
     =\frac{\bf{F}_n}{2\pi\rho R_n^2}
     +\sum\limits_{\substack{l=1\\l\neq n}}^N\left\{\frac{({\bf{p}}_n
  -{\bf p}_l)B_1}{d_{nl}^3}\right.\nonumber\\
  &&\qquad-\frac{R_l^2}{2d_{nl}^3}[R_nR_l\ddot{\bf p}_l+B_2\dot{\bf
     p}_l]\nonumber\\ 
  &&\qquad+\frac{3R_l^2}{2d_{nl}^5}({\bf p}_n-{\bf p}_l) 
    \left\langle ({\bf{p}}_n-{\bf p}_l)\cdot [R_nR_l\ddot{\bf
     p}_l+B_2\dot{\bf p}_l ]\right\rangle\bigg\}\,,\quad\label{eq:eq9}
\end{eqnarray}
where  overdots denote differentiation with respect to time, 
$B_1=R_nR_l^2\ddot{R}_l+2R_nR_l\dot{R}_l^2+\dot{R}_n\dot{R}_l
R_l^2$, $B_2=\dot{R}_nR_l+5R_n\dot{R}_l$, ${\bf p}_{l,n}$ are
the position vectors of the $l$th and $n$th bubble centres and
${\bf F}_n$ denotes the external force acting on the $n$th bubble.

Equation~(\ref{eq:eq8}) describes the radial oscillations of the $n$th
bubble in the cluster and Eq.~(\ref{eq:eq9}) governs its translational
motion. In these equations, similarly to Eq.~(\ref{eq:eq5}) and 
Eq.~(\ref{eq:eq6}) the pressure $P_n$ is defined as
\begin{eqnarray}
  P_n(R,\dot{R})&=&\left(P_0-P_v+\frac{2\sigma}{R_{n0}}\right) 
                    \left(\frac{R_{n0}}{R_n}\right)^{3\kappa}
                    -\frac{4\mu\dot{R}_n}{R_n}-\frac{2\sigma}{R_n}\nonumber\\
                &&-P_0-P_v-P_{ex}({\bf{p}}_n)\,,\label{eq:eq10}
\end{eqnarray}
where $P_{ex}({\bf p}_n)$ is the pressure of the driving ultrasound
wave in the centre of the $n$th bubble. The external
forces ${\bf F}_n$ are the sum of the primary Bjerknes force
\begin{equation}
  {\bf{F}}_{nB}=-\frac{4\pi}{3}R_n^3\nabla P_{ex}({\bf p}_n)
  \label{eq:eq11}
\end{equation}
and the force exerted on the bubble by the surrounding fluid (see
Ch.~8, Sec.~82 in \cite{Levich} and \cite{Lev49}), which in the case
of the oscillating bubble is given by \cite{Doi04}
\begin{equation}
  {\bf{F}}_{nL}=-12\pi\mu R_n\left (\dot{\bf p}_n - {\bf v}_{ex}({\bf p}_n) -  
         \sum\limits_{\substack{l=1\\l\neq n}}^N {\bf v}_{ln}\right)\,,   
  \label{eq:eq12}
\end{equation}
where ${\bf v}_{ex}({\bf p}_n)$ is the liquid velocity forced
by the driving pressure field in the centre of the $n$th bubble.
The fluid velocity generated by the $l$th bubble in the centre of the
$n$th bubble is given by 
\begin{eqnarray}
  {\bf v}_{nl}&=&\frac{R_l^2\dot{R}_l({\bf{p}}_n-{\bf p}_l)}{d_{nl}^3}\nonumber\\
              &&+\frac{R_l^3}{2d_{nl}^3}\left\{\frac{3({\bf p}_n
                 -{\bf p}_l)}{d_{nl}^2}
                  [\dot{\bf p}_l \cdot ({\bf p}_n-{\bf p}_l)]
                 -\dot{\bf p}_l\right\}\,.\quad\label{eq:eq13}
\end{eqnarray}
Note that although the model Eqs.~(\ref{eq:eq8})--(\ref{eq:eq13})
were derived for the case of microscopic gas bubbles driven by
high-pressure ultrasound fields \cite{Doi04}, all its equations
remain valid for mm-sized bubbles \cite{Levich}.

\begin{figure*}[t]
  \centerline{
    \includegraphics[width=16cm]{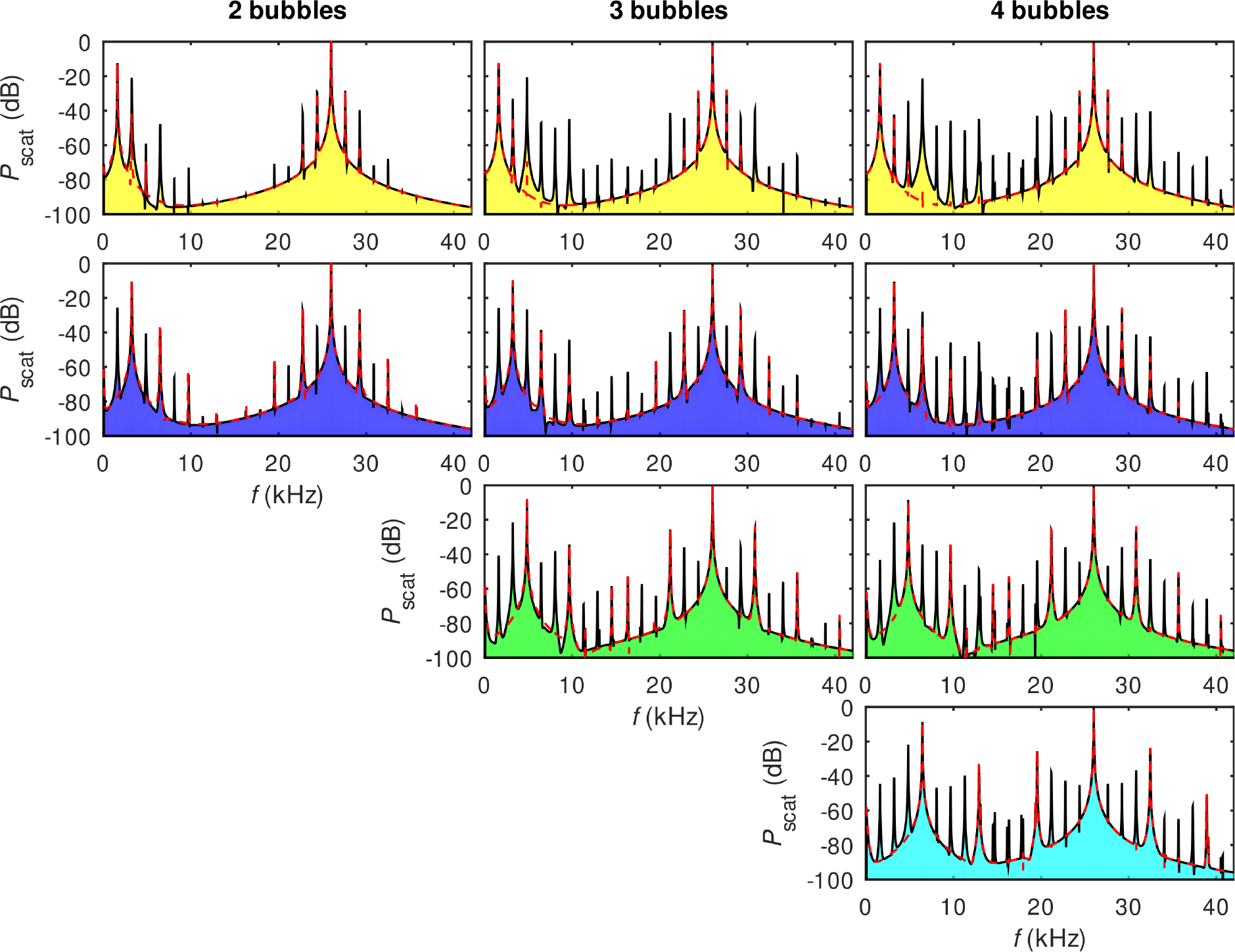}}
  \caption{Columns, from left to right, show the AFC spectra
    produced by the clusters consisting of 2, 3 and 4 gas bubbles with 
    the equilibrium radii $R_{n0}=1.95/n$\,mm, where $n$ is the
    number of bubbles in the cluster. The red dashed lines in each
    panel denote spectra of individual non-interacting stationary
    gas bubbles with the identical equilibrium radii. Computational
    parameters are given in the main text. Spacing between the
    spectral peaks is equal to the natural frequency of the largest
    bubble in a cluster. The number of peaks increases with the 
    bubble number in the cluster. Peaks corresponding to isolated gas
    bubbles are in good agreement with the ones in the spectra of
    bubble clusters. \label{fig:fig2}}
\end{figure*}

\section{Numerical results} \label{sec:Num}

Computations have been performed for the following fluid parameters
corresponding to water at $20^\circ$\,C: $\mu=10^{-3}$\,kg\,m/s,
$\sigma=7.25\times10^{-2}$\,N/m, $\rho=10^3$\,kg/m$^3$ and
$P_v=2330$\,Pa. We take the air pressure in a stationary bubble
to be $P_0=10^5$\,Pa and the polytropic exponent of air to be
$\kappa=4/3$.

The system of equations~(\ref{eq:eq8}) and (\ref{eq:eq9}) is solved
numerically using a fixed-step fourth order Runge-Kutta method
implemented in a customised subroutine \texttt{rk4} \cite{Mudrov}
ported from Pascal to Oberon-07 programming language. The accuracy
of this subroutine was tested by solving RP equation
(\ref{eq:eq5}) for a single gas bubble and comparing the result with
a solution obtained using a standard adaptive-step subroutine
\texttt{ode45} in the Octave software. While essentially the same
results were obtained using both subroutines, Oberon-07-based
computations are an order of magnitude faster and thus are
preferred for modelling multi-bubble clusters.    

We analyse an acoustic response of four clusters that 
consist of 2, 3 and 4 bubbles with the equilibrium radii 
$R_{n0}=R_{10}/n$, where $n$ is the number of the bubbles
and $R_{10}=1.95$\,mm is the typical bubble radius used
in \cite{Mak21}. The frequency of the driving
sinusoidal ultrasound wave propagating in the positive
$z$ direction is 26\,kHz and its peak pressure is 10\,kPa.
The coordinates of the centres of the bubbles are
$(x,y,z)=(-2.5R_{10},0,-2.5R_{10})$, $(2.5R_{10},0,2.5R_{10})$, 
$(0,4.33R_{10},0.2R_{10})$ and $(0,-4.33R_{10},-0.1R_{10})$.
This specific configuration resembles a typical bubble cluster
arrangement observed in our experiments \cite{Mak21}. 
Results qualitatively similar to those presented below
were obtained using other cluster configurations with the
same equilibrium radii of the bubbles and distances between them.
 
Figure~\ref{fig:fig2} shows the calculated spectra of the
bubble clusters (shown by colour-filled regions). Each
column shows the spectrum of the pressure scattered by the
individual bubbles within the cluster (calculated using
Eq.~(\ref{eq:eq7}) for each bubble in the cluster). We compare
these spectra with those of the isolated non-interacting stationary
bubbles of the same equilibrium radius (calculated using
Eq.~(\ref{eq:eq5}) and shown by the dashed lines).

Firstly, we notice that all the spectra exhibit the key features
pertinent to the generation of AFCs: the spectrum of the acoustic
response of each bubble consists of a series of well-defined and
equally spaced peaks. Secondly, we observe that the number of peaks
increases with the number of the bubbles in the cluster. The comparison
with the spectra of isolated non-interacting bubbles reveals that
the spectra of the bubbles in the clusters are composed of peaks
produced by all individual bubbles within the cluster. This is because
bubbles within a cluster are affected by pressure waves scattered by
their neighbours. Therefore, the spectrum of each bubble includes
additional peaks. 

For a two-bubble cluster, these observations agree
with predictions of our asymptotic analysis in Sec.~\ref{sec:Poincar}.
Indeed, in the leftmost column in Fig.~\ref{fig:fig2}) we can
identify (counting from from left to right) a pair of peaks
at the natural frequencies of the first and second bubble and
two other at the second harmonics of these frequencies. In each
pair, the peak corresponding to oscillations at the natural frequency
of the first (second) bubble has a larger magnitude than that
induced by the second (first) bubble in the cluster.
All these peaks give rise to the sideband structure around
the peak at the forcing frequency (26\,kHz), the highest peak in both
spectra (and also at its ultraharmonic at 42\,kHz that is not shown in
Fig.~\ref{fig:fig2}). The relative magnitude of the sideband peaks
follows the pattern of the peaks originating from oscillations at the
natural frequencies of the bubbles. The results of our numerical
simulation also indicate the presence of the third and fourth
harmonics of the natural frequencies in the spectrum of the cluster
response above the noise level. Capturing them analytically is
straightforward but would require retaining cubic and quartic terms in
$\epsilon$ in expansions (\ref{eq631}) thus leading to very long
algebraic expressions. For this reason, we do not present them here.

A similar but more complex picture is observed in the cases
of 3 and 4 bubbles, where the triplets and quadruples 
of the peaks corresponding to the natural frequencies of the 
bubbles and their harmonics can be identified. Furthermore,
these peak ensembles generate the sideband peak structures
around the forcing frequency, thereby forming a quasi-continuum of
equally-spaced peaks that is advantageous for AFC applications.

Of course, the magnitude of the peaks originating from the bubbles
of different sizes is different. However, whereas generating peaks
of the same magnitude would be advantageous for certain AFCs
applications, having peaks of different magnitude is, in general,
inconsequential as long as they are detectable and their frequencies
are stable (see \cite{Mak21} for a comprehensive discussion of this
technical aspect).
\begin{figure}[t]
  \centerline{\includegraphics[width=0.4\textwidth]{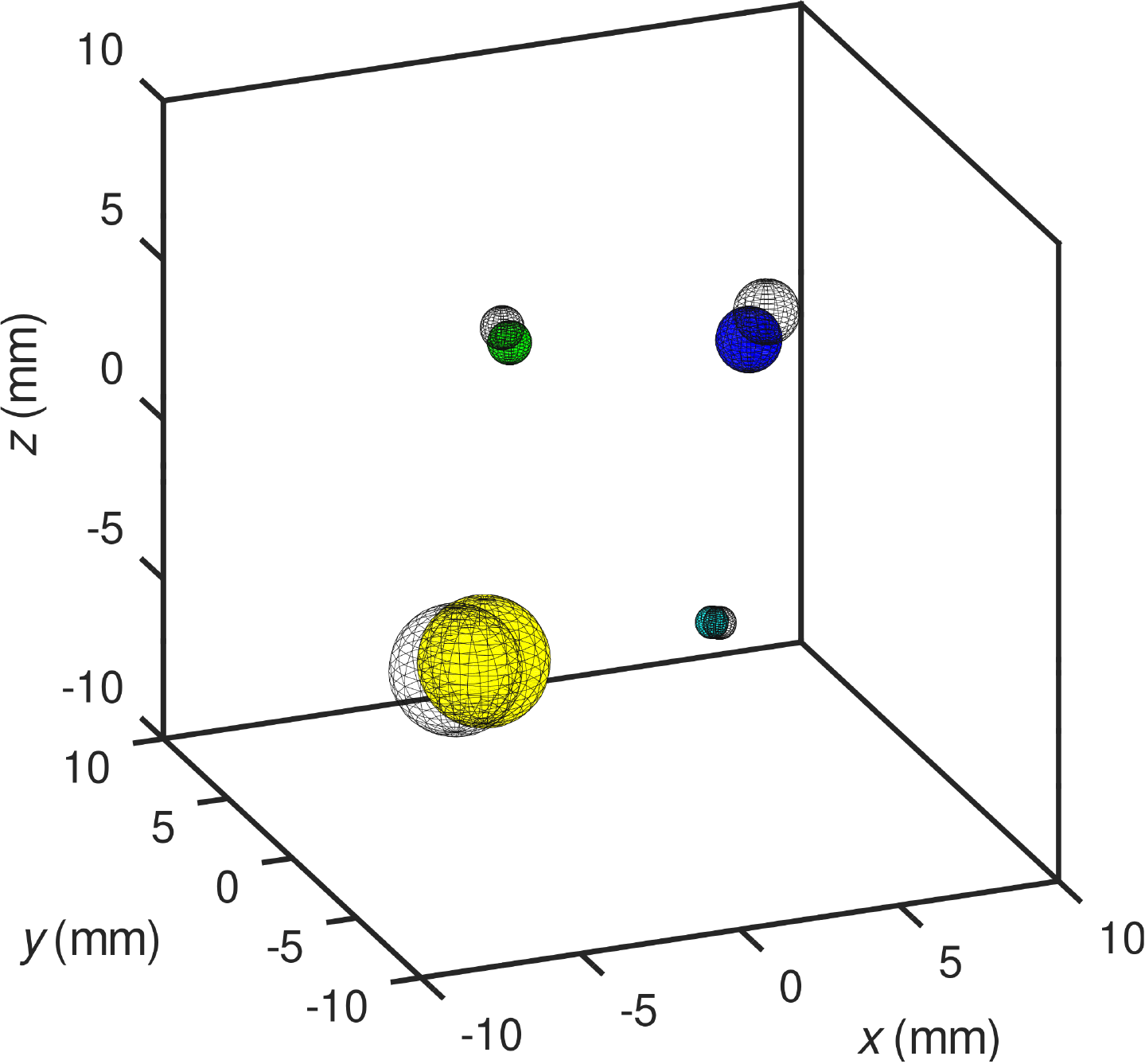}}
  \caption{Initial arrangement of bubbles in the four-bubble
    cluster (the transparent spheres) and the position of the same
    bubbles after 2000 periods of oscillations of the driving
    ultrasound wave (coloured spheres).\label{fig:fig3}}
\end{figure}

It follows from this discussion that accounting for the translational
motion of bubbles does not noticeably alter computational results. We
also note that the spectral line positions and shapes corresponding to
bubbles undergoing translational motion and interacting with their
neighbours are almost identical to those of isolated non-interacting
stationary gas bubbles (compare the colour-filled spectra with the
spectra shown by the red lines in each panel in Fig.~\ref{fig:fig2}).
This means that bubble translation does not affect their natural
frequency within the considered computational time. In support of this
conclusion, in Fig.~\ref{fig:fig3} we compare the initial spatial
bubble positions in the four-bubble cluster with the positions of the
same bubbles after 2000 periods of the driving ultrasound wave. 
In agreement with the results of the analysis presented in
Sec.~\ref{sec:Bjerknes} we observe that bubble attraction is weak so
that even after 2000 periods of oscillations the distance between
them remains sufficiently large for the cluster to generate a usable
AFC spectrum (and for the model to remain validity). 

\section{Conclusions}
We have proposed a new approach to the generation of spectrally-wide
AFCs using oscillations of polydisperse gas bubble clusters in
liquids. The plausibility of such an approach has been demonstrated
via theoretical analysis and numerical simulation. In the model
used in computations, we excite a bubble cluster with a low-pressure
ultrasound wave at a frequency that is higher than the natural frequency
of any bubble in the cluster. Bubbles within a cluster interact with
each other and their acoustic spectra are affected by their neighbours.
We choose the bubble sizes in such a way that their natural frequencies
become integer multiples of the natural frequency of the largest bubble
in the cluster. Because of that the spectra of individual bubbles contain
multiple peaks, each of which can be unambiguously associated with the
specific bubble within the cluster.

In agreement with the analysis and experimental observations reported
in our previous publication \cite{Mak21}, the interference of bubble
responses at their natural frequencies with the driving ultrasound wave
results in the amplitude modulation of the latter and in the appearance
of equally-spaced sideband peaks in the spectrum. Moreover, the
combination of sidebands and the ensemble of peaks originating
directly from the oscillations at the natural bubble frequencies
results in a quasi-continuum of equally spaced peaks. Its spectral
composition depends on bubble radii and the frequency of a driving
ultrasound wave. Therefore, it can be tuned by changing either of
these parameters. In particular, because mm-sized gas bubbles are
required to generate AFCs discussed in this work, a practical
realisation of the proposed approach is technically
straightforward. Indeed, a simple customised bubble generator
consisting of a standard air pump and a diffuser made of a
suitable porous material \cite{Mak21} would suffice to produce
bubble clusters with the configuration investigated in this
paper. We have also demonstrated that the attraction and potential
coalescence of oscillating bubbles due to the action of the secondary
Bjerknes force do not affect the generation of the AFC because
coalescence occurs at a timescale that is much larger than the 
time needed for a reliable recording of AFC signals. Finally,
our results are also expected to apply to smaller gas bubbles
driven by ultrasound waves in high kHz and MHz frequency ranges,
which paves the way for the generation of AFCs
suitable for a wide range of technologically important applications.

\begin{acknowledgments}
  ISM acknowledges the support from the Australian Research Council
    through the Future Fellowship (FT180100343) program.
\end{acknowledgments}


\providecommand{\noopsort}[1]{}\providecommand{\singleletter}[1]{#1}%

\end{document}